\pdfoutput=1
\documentclass[runningheads]{llncs}
\usepackage{graphicx}
\usepackage{pdfpages}
\usepackage{multirow}
\usepackage{adjustbox}
\usepackage{float}% http://ctan.org/pkg/float
\usepackage[ruled,vlined,linesnumbered]{algorithm2e}
\usepackage{hyperref}
\usepackage{breakcites}
\hypersetup{
    colorlinks=true,
    linkcolor=blue,
    urlcolor=blue,
    citecolor=blue,
}
\begin{document}
\title{A Robust and Accurate Approach to Detect Process Drifts from Event Streams}

%
%\titlerunning{Abbreviated paper title}
% If the paper title is too long for the running head, you can set
% an abbreviated paper title here
%
\author{Yang Lu\orcidID{0000-0002-9002-8650} \and
Qifan Chen\orcidID{0000-0003-1068-6408} \and
Simon Poon\orcidID{0000-0003-2726-9109}}
\authorrunning{Y. Lu et al.}
\titlerunning{Robust and Accurate Process Drift Detection}
% First names are abbreviated in the running head.
% If there are more than two authors, 'et al.' is used.
%
\institute{School of Computer Science, The University of Sydney, Sydney, NSW 2006, Australia 
\email{\{yalu8986, qche8411\}@uni.sydney.edu.au}\\
\email{simon.poon@sydney.edu.au}}
\maketitle              % typeset the header of the contribution

\begin{abstract}
Business processes are bound to evolve as a form of adaption to changes, and such changes are referred as process drifts. Current process drift detection methods perform well on clean event log data, but the performance can be tremendously affected by noise. A good process drift detection method should be accurate, fast, and robust to noise. In this paper, we propose an offline process drift detection method which identifies each newly observed behaviour as a candidate drift point and checks if the new behaviour can signify significant changes to the original process behaviours. In addition, a bidirectional search method is proposed to accurately locate both the adding and removing of behaviours. The proposed method can accurately detect drift points from event logs and is robust to noise. Both artificial and real-life event logs are used to evaluate our method. Results show that our method can consistently report accurate process drift time while maintaining a reasonably fast detection speed.

\end{abstract}
\section{Introduction}

Business processes are continuously evolving in order to adapt to changes. Changes are often responses to different factors which can be planned or unexpected. For example, a planned change can be caused by the introduction of a new regulation, and an unexpected change can be caused by an emergency (eg. the COVID-19 outbreak). In the field of process science, such changes are called process drifts.

It has been argued that assuming a process model to be stable is unrealistic~\cite{Bose2014}. It is important for us to detect process drifts as accurately as possible. On the one hand, unexpected changes can cause losses to organizations. Detecting such drifts can help us make appropriate responses to changes. On the other hand, most current algorithms to discover process models assume the process to be in a steady-state and ignore process drifts~\cite{wmp2016}. Detecting and understanding process drifts can help us to understand the evolving nature of processes.

Statistically, a process drift point is a time point when there is a significant difference among the process behaviours before and after the drift point~\cite{Maaradji2017,Ostovar2016,Maaradji2015}. Various process drift detection methods have been proposed in the last decade~\cite{Bose2014,zheng2017,Maaradji2017, Seeliger2017,Martjushev2016,Stertz2018,Stertz2019,Brockhoff2020,Ostovar2016,Hompes2017,Liu2018,Lin2020,Yeshchenko2019,Carmona2012,Accorsi2012,Bose2011,Maaradji2015}. However, many of these methods assume the input event log data to be clean, and their abilities to handle noise can vary. In~\cite{wmp2016}, noise is defined as "the event log contains rare and infrequent behaviours not representative the typical behaviour of the process". Such behaviours are infrequent and cannot cause a significant change to the process behaviours. For example, if an activity is skipped only once in one process execution record, it is more likely to be an infrequent behaviour instead of a process change. However, noise is known to have big impacts on process drift detection accuracy.

In this paper, we consider a process drift as either the adding or removing of behaviours which can signify significant changes to the behaviours of the original process. We focus on offline process drift detection from the control-flow perspective. We propose an event-stream based process drift detection method which is accurate, robust to noise and reasonably fast. When a new behaviour is observed in the event log, we treat it as a candidate drift point and verify if it can signify significant changes to the current process through statistical tests. Both artificial and real-life event logs are used to evaluate the method.

The rest of this paper is structured as follows: Section \ref{Section:2} is a literature review of related work. Section \ref{Section:3} introduces formal definitions of some terms. Section \ref{Section:4} introduces the proposed method. The evaluation results are presented in Section \ref{Section:5} and Section \ref{Section:6}. We finally conclude the paper in Section \ref{Section:7}.

\section{Background}
\label{Section:2}
\subsection{Detecting Process Drifts by Statistical Tests}
%Different methods to detect process drifts have been proposed in the last decade~\cite{Bose2014,zheng2017,Maaradji2017,Bolt2016,Martjushev2016,Stertz2018,Stertz2019,Brockhoff2020,Ostovar2016,Hompes2017,Liu2018,Lin2020,Yeshchenko2019,Carmona2012,Accorsi2012,Bose2011,Maaradji2015}. 
A general approach to detect process drifts is to use a sliding window to obtain two consecutive samples in the event log, naming as reference and detection windows. The two windows are then moving through the event log trace by trace or event by event. Then samples within each window are transformed into a set of features, and if statistical hypothesis tests show that there is a significant difference before/after a certain time point among these features, a drift is reported. 

Early approaches such as~\cite{Bose2014,Bose2011} extract features to represent each sample of the event log. Then statistical hypothesis tests are applied to detect process drifts among feature vectors. Based on~\cite{Bose2014},~\cite{Martjushev2016} applies adaptive window approaches to automatically adjust window sizes. Those methods require users to select features to be used, which require background knowledge about the drifts in input event logs.

The ProDrift run-based method~\cite{Maaradji2017,Maaradji2015} transforms each trace into a partial-order run which is a graph representation of a trace eliminating the order between parallel events. Then chi-square tests are applied to detect if there are any significant changes among the distribution of partial-order runs between two consecutive windows. The method is fully automated with the capability to categorize certain drift types. In addition,~\cite{Maaradji2017,Maaradji2015} also aim at eliminating the impact of noise by performing a number of consecutive tests. However, since each trace is only counted once in each window, the samples used for statistical tests are relatively small, returning unreliable results especially when input logs have high variability (eg. when the event log contains noise). The ProDrift event-based method~\cite{Ostovar2016} improves~\cite{Maaradji2015} by treating event logs as event streams and using the count of alpha+ relations\footnote{Alpha+ relations define a set of relations between activities which are conflict, concurrency, causality, length-one loop and length-two loop. For their formal definitions, please refer to \cite{Alves2004}.} as features for statistical tests. On the one hand, process drifts during the execution of traces can be detected. On the other hand, since the number of alpha+ relations is much larger than traces in each window, the statistical tests in~\cite{Ostovar2016} are more reliable. In addition, the ProDrift event-based method~\cite{Ostovar2016} can also filter out infrequent behaviours and can work both in online and offline settings. It is also the basis of the new approach to characterize process drifts in~\cite{Ostovar2020}. The ProDrift event-based method~\cite{Ostovar2016} requires parameters such as noise filtering thresholds from users.
%However, it suffers from relatively longer delays (i.e. the distances from the actual process drift points to the reported drift points) due to the nature of statistical tests.% 

When using statistical tests to detect process drifts in event logs, the distances between the actual process drift points and the reported drift points are relatively longer, resulting in lower detection accuracy.
\subsection{Other Process Drift Detection Methods}
To improve detection accuracy, the TPCDD method~\cite{zheng2017} and the LCDD method~\cite{Lin2020} are proposed. Both methods can achieve high accuracy. The TPCDD method~\cite{zheng2017} firstly transforms the whole event log into a relation matrix, and whenever a new behaviour is detected or an existing behaviour is removed, if it lasts for a certain period, a new drift point is reported. The LCDD method~\cite{Lin2020} firstly finds a time window where the sub-log within the window is locally complete. Then whenever a new behaviour is observed or an existing behaviour is removed, a drift point is reported. Although these two methods can return highly accurate results on artificial logs, they are very sensitive to noise.

Other methods are also proposed to detect process drift points.~\cite{Accorsi2012} detects process drifts based on the change of distances between each pair of activities. Loops and parallel behaviours are ignored, resulting in possible failures.~\cite{Carmona2012} abstracts initial traces into a polyhedron and checks if subsequent traces are within the polyhedron, a drift is detected if a trace is outside the polyhedron.~\cite{Carmona2012} is the first concept drift detection method which can be used in online settings, but it suffers from long execution time.

Instead of focusing on detection accuracy, some methods focus more on understanding how the process model evolved over time.~\cite{Seeliger2017} mines process models for different time periods and compares graph matrices of different models.~\cite{Stertz2018,Liu2018} mine models for the first period of time and perform conformance check on each new trace. A drift point is reported if there is a significant change on the conformance checking results.~\cite{Yeshchenko2019} applies Declare miners to represent the process, and a comprehensive visualisation is provided to understand process drifts. These methods provide comprehensive analyses of process drifts, but usually suffer from relatively longer execution time and lower accuracy. 

Some methods also focus on process drifts from other perspectives other than the control-flow perspective. For example,~\cite{Stertz2019} detects process drifts from the data value perspective (eg. the change of activity attribute values),~\cite{Brockhoff2020} applies the earth mover's distance to detect time and control-flow drifts together (eg. the change of activity execution time).

In summary, existing methods which are highly accurate are sensitive to noise in event logs. Methods which are capable of handling noise could improve their accuracy. A new method which requires fewer user-inputs and can produce highly accurate results while properly handling noise is needed in this field.
\section{Preliminaries}
\label{Section:3}
\begin{definition}[Process drift point~\cite{Maaradji2017,Ostovar2016,Maaradji2015}]
A process drift point is a time point when there is a statistically significant difference among the observed process behaviours before and after the time point. 
\end{definition}

\begin{definition}[Event log, Trace, Activity, Event]
An event log L is a multiset of traces where each trace $t_i$ is a sequence of events in a set E, i.e. E = $\{e_1, e_2, ......, e_n\}$, and each event corresponds to a single activity A.
%Let N be a process model and A be the set of activities within N. An event log $L \subset A^*$ of process model N is a collection of traces corresponding to the execution of N. $t \in L$ is a trace of log L.
\end{definition}

\begin{definition}[Directly-follows relation]
Let L be an event log of a process model N, let A, B be two activities in L. Then there is a directly-follows relation from A to B, denoted as $A >_L B$, if there exists a trace $t \in L$ where $t = <......, A, B, ......>$. 	
\end{definition}
\section{Concept Drift Detection}
\label{Section:4}
\begin{figure}
\includegraphics[width=\textwidth]{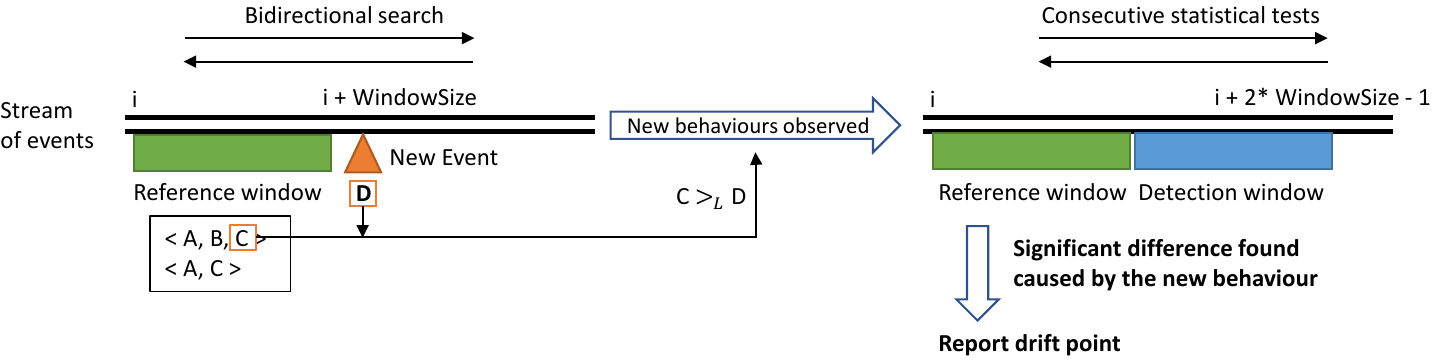}
\caption{Overview of the proposed method} \label{approach_summary}
\end{figure}

Fig.\ref{approach_summary} shows an overview of our proposed method. Our method firstly converts the input event log into a stream of events where events are indexed and ordered by their timestamps. Then a reference window is built and continuously moves through the event stream. A sub-log is built including all events within the reference window. Each time the reference window moves, the sub-log is updated and the event immediately follows the reference window is peeked. If the peeked event brings a new behaviour which cannot be observed in the sub-log corresponding to the reference window, we treat it as a candidate drift point and check if the new behaviour can signify a significant difference to the original behaviours of the process through statistical tests. If so, a drift point is reported.

\subsection{Selection of Features}
\label{Section:4.1}
The first step of designing a process drift detection method is to find a feature which can represent the behaviours of the process, and changes of such features should reflect changes in process control-flow structures. As the proposed method relies on a single event to determine possible process drifts, we decide to use directly-follows relations as features to represent process behaviours for two reasons: 1) Most current process discovery algorithms are based on directly-follows relations~\cite{wmp2016}, changes in process control-flow structures are highly likely to result in changes of directly-follows relations. 2) By peeking one event after the reference window, a directly-follows relation could be obtained. It is worth mentioning that alpha+ relations used by~\cite{Ostovar2016} are not suitable for our method as an alpha+ relation cannot be determined by a single peeked event. 

\subsection{Validation of Candidate Drift Points}
\label{Section:4.2}
Observing a new directly-follows relation means a possible process drift is detected. However, it could also be noise inside the event log. Whenever a new directly-follows relation is observed, we treat it as a candidate drift point. Statistically, a process drift point should be treated as a time point $t$, and there is a significant difference between process behaviours before and after $t$ \cite{Maaradji2017,Ostovar2016,Maaradji2015}. Although noise can bring new observed directly-follows relations in event logs, significant changes to the process behaviours will not be signified.

To confirm if a candidate drift point is an actual drift point, statistical tests are applied to check if a significant difference is caused. Firstly, a detection window is built after the reference window, and a contingency matrix is built to report the frequencies of each type of directly-follows relations in both the reference and detection windows. Then the G-test of independence\footnote{the G-test of independence is a non-parametric statistical hypothesis test.}\cite{Woolf1957} is applied and a P-value is returned. If the P-value is less than a certain threshold, which is typically 0.05, we conclude there is a significant difference between process behaviours before and after the candidate drift point.

If the G-test of independence shows there is a significant difference between process behaviours before and after a candidate drift point, it is likely to be an actual drift point. However, if the candidate drift point is close to an actual drift point, the low P-value could be caused by other directly-follows relations instead of the new observed one.

To check if the new observed directly-follows relation contributes to the low P-value, the adjusted standardized residual (ASR) of the new directly-follows relation in the detection window is calculated. If $ASR > 1.96$, we conclude the number of the new observed directly-follows relation is significantly larger and is an influential point to the test score. For details about this, we refer to~\cite{Agresti2003}.

Similar to previous studies such as~\cite{Maaradji2017,Ostovar2016,Maaradji2015}, a number of consecutive statistical tests are performed before a conclusion can be made to avoid sporadic stochastic oscillations. For details, please refer to Algorithm \ref{algorithm:1}, lines 9 -20.

\subsection{Bidirectional Searches}
\label{Section:4.3}
\begin{figure}
\includegraphics[width=\textwidth]{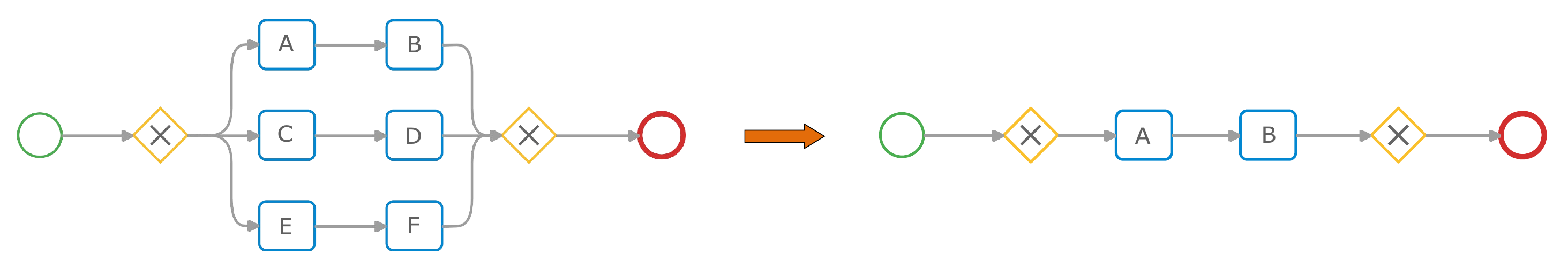}

\caption{An example process drift from model 1 (left) to model 2 (right)} \label{remove_example}
\end{figure}

A change in process models can cause both the adding and removing of directly-follows relations. Detecting an added directly-follows relation can be simply done by checking if the newly observed directly-follows relation exists in the reference window. However, the removing of directly-follows relations cannot be detected by peeking one event immediately after the reference window. A possible solution is to build a detection window immediately after the reference window and checks if any directly-follows relations are removed. Such a method can affect detection accuracy. Fig.\ref{remove_example} shows an example process drift. Suppose model 1 is shifted into model 2 at time $t$. $E >_L F$ and $C >_L D$ will no longer be observed after $t$. However, suppose the last appearance of $E >_L F$ is at $t_1$ which is earlier than $t$, $t_1$ could be treated as the drift point by mistake, reducing the detection accuracy.

To solve the problem, we perform both forward and backward searches on the event stream to detect process drifts. When performing backward searches, the removing of directly-follows relations is shown as the adding of directly-follows relations. There are two advantages of performing bidirectional searches: 1) When a process drift causes both adding and removing of directly-follows relations, if the drift is missed by one search, there is one more chance for it to be detected in another search. 2) the accuracy of detection can be improved. 

It is worth mentioning that performing bidirectional searches will not double the amount of time required to detect process drifts. Each time a G-test is performed, its resulting P-value will be stored and if another G-test is required at the same position, the P-value can be retrieved within constant time. Furthermore, each time when a G-test is performed, the ASR of each directly-follows relation can also be computed and stored. As a result, no duplicate statistical tests will be performed. We show that our algorithm is efficient to detect process drifts in Section \ref{Section:5} and Section \ref{Section:6}.

\subsection{The Framework of the Proposed Method}
\label{Section:4.4}
Finally, we present the forward detection method in Algorithm \ref{algorithm:1}. Since the same approach applies to the backward detection but in a reverse direction, we do not present it separately. Lines 1 - 8 build a reference window, peek the next event and see if a new directly-follows relation is found (Section \ref{Section:4.1}). Whenever a new directly-follows relation is observed, the new event is treated as a candidate process drift point. Lines 9 - 25 perform statistical tests to confirm if a candidate process drift point is an actual drift point (Section \ref{Section:4.2}).

\subsubsection{When a noise is close to the real process drift point}
A challenge to the proposed method is when a new directly-follows relation is observed which is noise but is close to the actual process drift point. Although the problem can be solved by calculating ASRs, it fails to solve the case when the noisy directly-follows relation is the same as one of the added directly-follows relations after the actual drift point. For example, suppose directly-follows relations $A >_L B$ and $C >_L D$ are added to the process after a process drift at time $t$, if a noisy directly-follows relation $A >_L B$ is observed at time $t_0$ which is earlier than $t$, $t_0$ could be treated as a drift point by mistake. To overcome this issue, two measures are taken: 1) When performing a number of consecutive tests, we not only move the windows forward but also move the windows backward (Algorithm \ref{algorithm:1}, lines 9 - 20). 2) By moving the window backward, the noisy $A >_L B$ could be differentiated. However, since it is close to the real process drift point, having $A >_L B$ in the reference window can avoid $A >_L B$ from being observed as a new behaviour when arriving at the real drift point. As a result, if a new observed directly-follows relation fails statistical tests, we delete it from the reference window (Algorithm \ref{algorithm:1}, line 28).

\begin{algorithm}[]
\label{algorithm:1}
\SetAlgoLined
\KwIn{eventStream, windowSize, numOfConsecutiveTests}
refStartPosition $\gets$ 0\\ \tcp{The index of the first event in the reference window} 
refSubLog $\gets$ getSubLog(eventStream, windowSize, refStartPosition)\\
refDfRelations $\gets$ getDfRelations(refSubLog)\\ \tcp{The directly-follows relations of the sub-log}

%findDirectlyFollowsRelations(refSubLog)\;

 \While{$refStartPosition + 2 \cdot windowSize + numOfConsecutiveTests< eventStreamSize$}{
  e $\gets$ getNewEvent()\\ \tcp{Peek the first event immediately after the reference window}
  $>_e \gets getNewDfRelation(e, refSubLog)$\\  \tcp{Get the new directly-follows relation brought by e}
  numOfSatisfiedTests $\gets$ 0\\
  \If{$>_e \neq null$ AND $>_e \notin refDfRelations$}{\tcp{A candidate drift point is found}
    
    \For{$i\gets0$ \KwTo $2 \cdot numOfConsecutiveTests$}{
    refTestSubLog $\gets$ Sub-log for window starting with event refStartPosition - numOfConsecutiveTests + i\\
    decTestSubLog $\gets$ Sub-log for window starting with event refStartPosition - numOfConsecutiveTests + windowSize + i\\
    Compute Contigency matrix based on the frequency of directly-follows relations in refTestSubLog and decTestSubLog\\
    Perform G-test on the matrix and get pValue\\
        \If{pValue is smaller than the threshold}{
            Compute $ASR$ for $>_e$\\
            \If{ASR is significant}{
                numOfSatisfiedTests ++\\
            }
        }
      
    }

   }
   \eIf{$numOfSatisfiedTests = 2 \cdot numOfConsecutiveTests$}{
        Report drift point e\\
        RefStartPosition $\gets$ index of e\\
        Update refSubLog and refDfRelations\\\tcp{Move the beginning of the reference window to the new detected change point}
        
   }{
        \If{$>_e \neq null$ AND $>_e \notin refSubLog$}{Remove $>_e$ from refDfRelations}
        RefStartPosition ++\\
        Update refSubLog and refDfRelations\\ \tcp{Move the reference window by one event}
   }
  
 }
 \caption{Forward Detection}
\end{algorithm}

\section{Evaluation on Synthetic Data}
\label{Section:5}
The proposed method is implemented as a stand-alone Java application. All the code, data and results are publicly-available\footnote{\url{https://github.com/bearlu1996/ProcessDrifts}}.
\subsection{Evaluation Design}
We firstly collect the 72 artificial event logs from~\cite{Maaradji2015} which are generated from an artificial process model containing 1 start event, 3 end events, 8 gateways and 15 activities.~\cite{Maaradji2015} systemically alters the base model by 12 simple patterns shown in Table \ref{patterns}. Each simple change pattern can be categorized as Insertion(I), Resequentialization(R) and Optionaliztion(O). Then the base model is also altered according to 6 composite change patterns (RIO, ROI, IOR, IRO, OIR, ORI). For each change pattern, 4 logs with 2500, 5000, 7500, 10000 traces are generated with a sudden process drift after every 10\% of traces\footnote{9 drift points are included in each log.}. The synthetic data set has also been used to evaluate drift detection algorithms in~\cite{Maaradji2017,Lin2020}.

Since branching frequency changes cannot be reflected on changes of directly-follows relations, it cannot be detected by our proposed method. We exclude its corresponding four event logs from our evaluation. Then for each event log, we insert noise by randomly adding and removing 10\%, 20\% and 30\% of events. To avoid biases caused by randomly generated noise, we generate 10 logs for each log at each noise level and take the average results in the following parts. In total, we evaluate our algorithm on 4148 synthetic event logs\footnote{$4148 = 68 \times 6 \times 10 + 68$} including logs without noise. It has to be noted that inserting noise will not change the trace indexes of process drifts\footnote{We do not add/remove traces into/from the event logs}.

In this section, when two drift points are reported by searches from different directions, and the distance between them is smaller than the window size, we only take the point with a smaller index.

Finally, since the drifts in the artificial event logs are inter-trace drifts (i.e. drifts occur between complete trace executions), we stream the events in the order from the first event in the first trace to the last event in the last trace\footnote{For example, event 0 refers to the 1st event in the 1st trace, event 1 refers to the 2nd event in the 1st trace ... the last event refers to the last event in the last trace.} in both our method and the baseline so that these drifts can also be detected by event-stream based algorithms. In this section, trace ids are used to represent the location of all process drifts\footnote{When an event id is reported, we refer to the id of its corresponding trace.}. 

\begin{table}[H]
\centering
\caption{Simple control-flow change patterns.} \label{patterns}
\includegraphics[width=\textwidth/2]{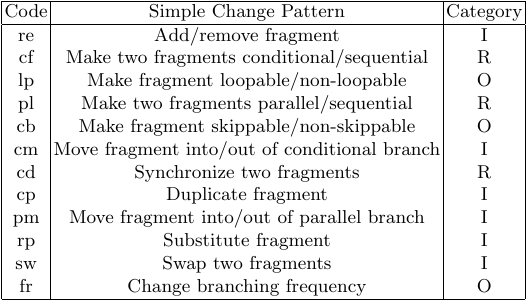}

\end{table}
%\begin{table}[H]
%
%\caption{Simple control-flow change patterns}
%\label{patterns}
%\begin{tabular}{ |c|c|c|c| } 
%\hline
%Code & Simple Change Pattern & Category \\
%\hline
%re& Add/remove fragment & I \\ 
%cf& Make two fragments conditional/sequential & R \\ 
%lp& Make fragment loopable/non-loopable & O \\ 
%pl& Make two fragments parallel/sequential & R \\ 
%cb& Make fragment skippable/non-skippable & O \\ 
%cm& Move fragment into/out of conditional branch & I \\ 
%cd& Synchronize two fragments & R \\ 
%cp& Duplicate fragment & I \\ 
%pm& Move fragment into/out of parallel branch & I \\ 
%rp& Substitute fragment & I \\ 
%sw& Swap two fragments & I \\ 
%fr& Change branching frequency & O \\ 
%
%\hline
%
%\end{tabular}
%\end{table}

\subsubsection{Evaluation Metrics}
Standard f-score metric for evaluating process drifts detection approaches is used in our evaluation~\cite{zheng2017,Lin2020} where $precision = TP / (TP + FP), recall = TP / (TP + FN), f-score = 2 * precision * recall / (precision + recall)$. TP refers to true positive, FN refers to false negative, and FP refers to false positive. To describe the three variables, an error tolerance (ET) is defined. A TP is detected if a change point t is detected where the actual drift point is within the integer interval [t - ET, t + ET]. A FP is detected if a change point t is detected and there is not an actual drift point within the integer interval [t - ET, t + ET]. A FN is detected if an actual process drift exists in the integer interval [t - ET, t + ET] where no change points are detected.
\subsubsection{Baseline}
Among several existing methods, the ProDrift event-based method~\cite{Ostovar2016} seem to be more capable in handling noise than other popular methods such as~\cite{zheng2017,Maaradji2017,Lin2020,Maaradji2015}. As suggested by its documentation, we change the noise filtering threshold to 0 and drift sensitivity to "very high" for noise-free logs, and we use its default settings for all other logs (adaptive window is enabled for all tests).
%As stated before, very few process drift detection approaches have the capability of handling noises in event logs. We find that the ProDrift event-based~\cite{Ostovar2016} is the best approach which can handle noises. Other popular algorithms such as~\cite{zheng2017,Maaradji2017,Lin2020} obtain very low f-scores with noisy event logs. 

\subsection{Evaluation on Different Parameter Settings}
\begin{figure}
\includegraphics[width=\textwidth]{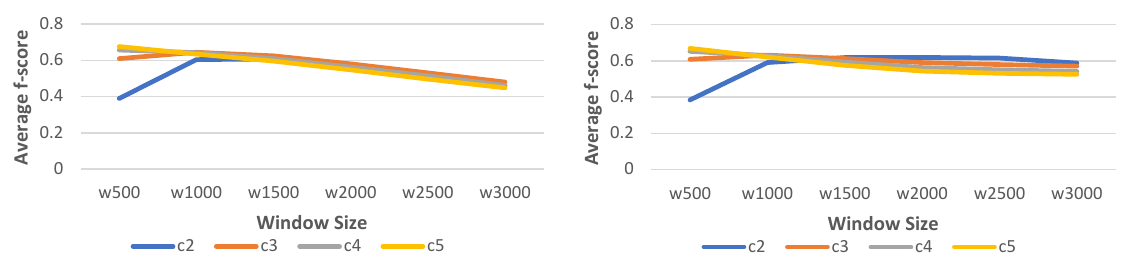}

\caption{Average f-scores (ET = 10) under different settings including logs of size 2.5k (left) and excluding logs of size 2.5k (right). Each color represents a setting of number of consecutive tests. For example, c2 means WindowSize / 2 consecutive tests required in one direction (Section \ref{Section:4.4}).} \label{parameters_10}
\end{figure}

In the first experiment, we evaluate the impact of window sizes and the number of consecutive tests on the detection results. We test a total of 6 different window sizes ranging from 500 to 3000, and we combine them with 4 different number of consecutive tests, ranging from windowSize / 5 to windowSize / 2. For each of the 24 settings, we run all the synthetic event logs and calculate the average f-scores. Fig.\ref{parameters_10} shows the average f-scores when the error tolerance is set to 10. 

Overall, the impacts of the number of consecutive tests to f-scores are small. When the number of consecutive tests is set to be WindowSize / 2, the accuracy is slightly higher and more consistent in most cases unless a small window size is set. We decide to set the number of consecutive tests to WindowSize / 2 as the default setting for our method.

With the number of consecutive tests being empirically set, the only user input required is the windowSize, When logs of size 2.5k are included (Fig.\ref{parameters_10} left), the average f-Scores drops after the window size of 1000 since the window size becomes larger than the distance between two consecutive process drift points. We then remove logs of size 2.5k from the calculation (Fig.\ref{parameters_10} right), results show that f-Scores are less sensitive to the choice of window sizes. Although a larger sample can result in more reliable statistical test results, having a larger window size could increase the chance of treating a new observed noisy directly-follows relation as a real drift point when it is close to the actual drift point, and the noisy directly-follows relation is the same as one of the added directly-follows relations after the drift point (Section \ref{Section:4.4}). It is worth mentioning that the average f-score of the baseline is only 0.03 when ET = 10, which is much lower than our method. We also calculate average f-scores when ET = 50 and obtain similar observations. 

It is also noticed that the choice of window sizes is related to the distance between two consecutive process drifts. For most current window-based process drift detection approaches, the accuracy drops when the window size is larger than the minimal distance between two consecutive drifts, or the window size is too small that event logs within windows are incomplete even if adaptive window approaches are implemented. In the remaining text, we report the results with the number of consecutive tests = WindowSize / 2. In section \ref{Section:5.3} and \ref{Section:5.4}, we set the window size of our method to 1500.

%\begin{figure}
%\includegraphics[width=\textwidth]{parameters_50.pdf}

%\caption{Average f-scores (ET = 50) under different settings including logs of size 2.5k (left) and excluding logs of size 2.5k (right).} \label{parameters_50}
%\end{figure}

\subsection{Comparing with the Baseline on Different Change Patterns}
\label{Section:5.3}
In the second experiment, the accuracy of our proposed method and the baseline is compared for each change pattern and under different noise levels.

\begin{figure}
\includegraphics[width=\textwidth]{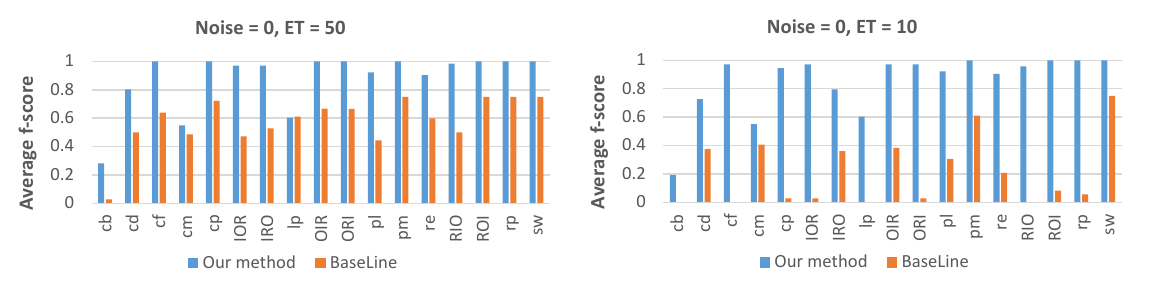}

\caption{Average f-scores per change pattern for noise-free logs comparing to the baseline with ET = 50 and 10.} \label{noise-free}
\end{figure}

We firstly run both methods on the 68 noise-free event logs, and the results are presented in Fig.\ref{noise-free} where each f-score is averaged over 4 logs with different sizes. When ET = 50, our method achieves an average f-score of 0.88 while the baseline achieves 0.58. When ET = 10, our average f-score achieves 0.85, which is close to the results when ET = 50. However, the baseline drops to 0.21, which means our method is more accurate. 

\begin{figure}

\includegraphics[width=\textwidth]{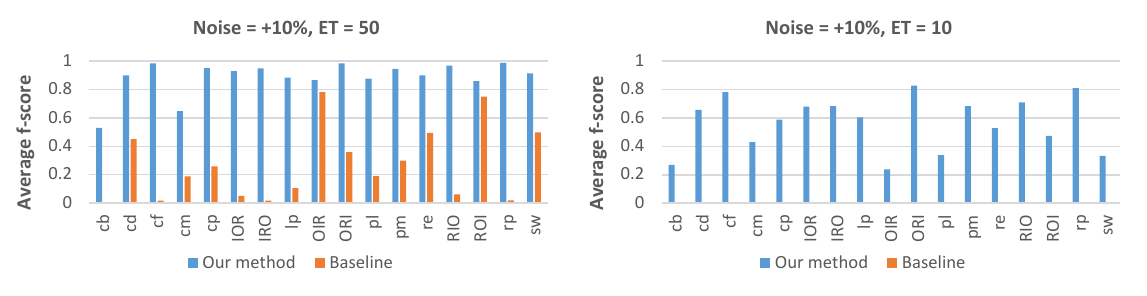}
\includegraphics[width=\textwidth]{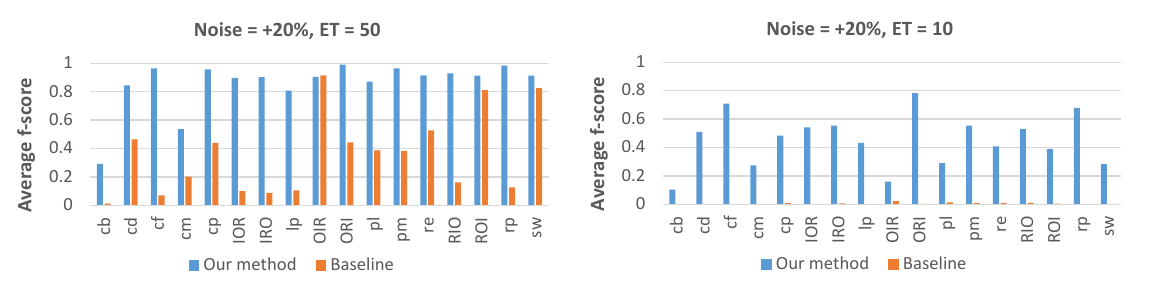}
\includegraphics[width=\textwidth]{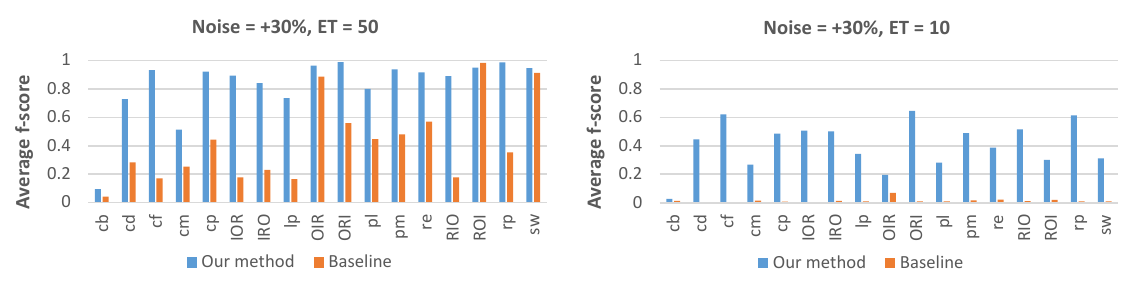}
\includegraphics[width=\textwidth]{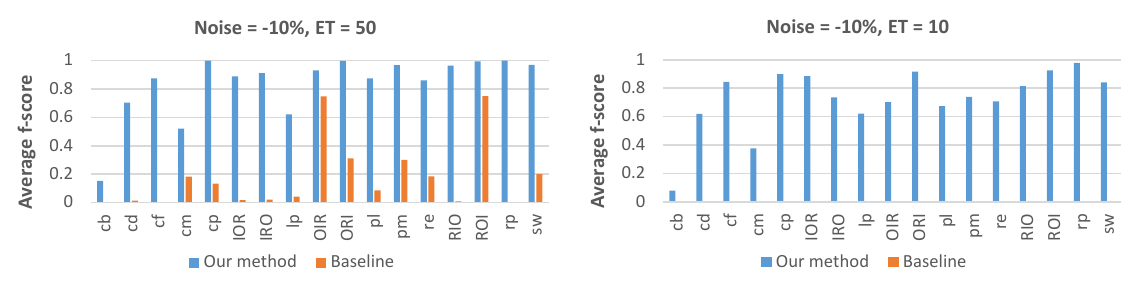}
\includegraphics[width=\textwidth]{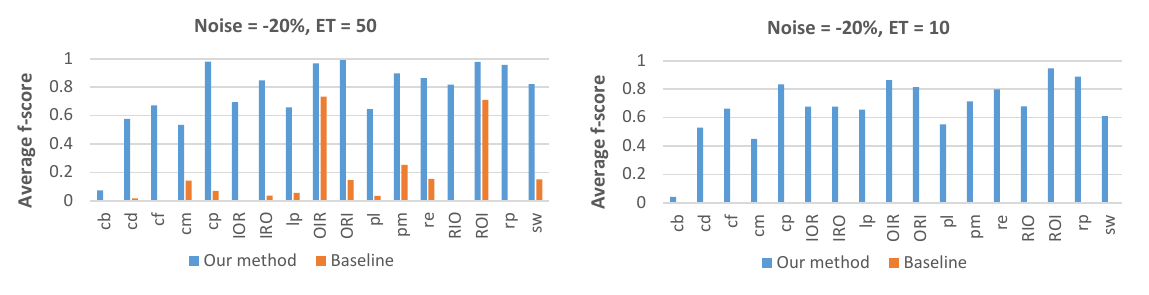}
\includegraphics[width=\textwidth]{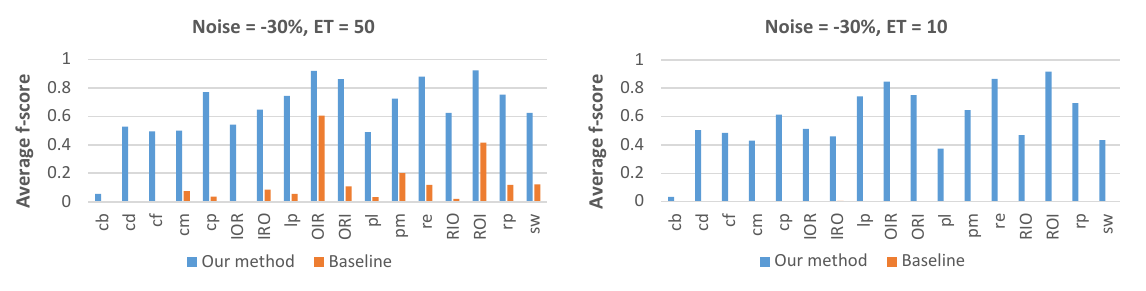}
\caption{Average f-scores per change pattern comparing to the baseline with ET = 50 and 10. For noise levels, + refers to inserting activities, and - refers to removing activities.} \label{noise}

\end{figure}

Fig.\ref{noise} shows the average f-scores for each change pattern under different noise levels when ET = 50 and ET = 10 where each f-score is averaged over 40 logs. When ET = 50, our method achieves an average f-score\footnote{Among all the logs with noise.} of 0.8 and an average of 0.57 when ET = 10. Comparing to the baseline, our method wins in almost all cases. When noise is inserted into the log, the baseline can achieve satisfied f-scores for a few change patterns when ET = 50. However, when ET = 10, the f-scores of the baseline drops dramatically, of which the f-scores are 0 in most cases.

It is also interesting to find that our method performs better when removing events from the event log. The main reason is that when inserting events into logs, the probability that a noisy directly-follows relation is inserted before a drift point which is the same as one of the added directly-follows relation after the drift point is higher (Section \ref{Section:4.4}). Thus, our method could report process drifts earlier than the actual drift points, causing lower f-scores when ET is small. We find that this is the biggest factor affecting the results. 

%Another finding is that for some change patterns, the baseline performs better when more events are added. This could be caused by three reasons: 1) Same as stated before, when noise behaviours are the same as new behaviours but appear earlier, drifts can be reported earlier. Since the baseline suffers from long detection delay, reporting results earlier could reduce the delay and have a chance to obtain a higher f-score. 2) If inserted behaviours are the same as new added behaviours and are inserted in correct positions, advantages are provided for statistical tests, making them easier to find a significant change 3) The results could be sensitive to user-defined parameters under different noise levels.
Finally, we also calculate the average precision among both methods when ET = 50. Our method achieves an average precision of 0.97 among all event logs while the precision of the baseline is only 0.3. A high precision indicates that our method will not return too many results which are not actual process drifts or mistakenly treat infrequent behaviours as process drift points, saving the time it takes to validate each drift point. In conclusion, our method is more accurate and reliable than the baseline for both event logs with or without noise.

\subsection{Execution Time}
\label{Section:5.4}
In the last experiment, we run both our method and the baseline on all artificial logs and record their execution time. The platform is equipped with Intel i7-9700 (8 cores, 8 threads) and 32GB RAM, running Windows 10 (64 bit) with a heap space of 16GB. Among the 4148 event logs, our method takes 0.03ms (min 0.01ms, max 0.13ms) for each event on average while the baseline takes 0.1ms (min 0.05ms, max 0.26ms) where average execution time for each event = total execution time / number of events. The results indicate that our method can detect process drifts efficiently and can be potentially applied in online settings.

\section{Evaluation on Real-life Data}
\label{Section:6}
We evaluate our algorithm on the BPI Challenge 2020 (BPIC2020) data-sets\footnote{\url{https://icpmconference.org/2020/bpi-challenge/}}. The BPIC 2020 data-sets collect a total of five event logs of travel reimbursement processes at Eindhoven University of Technology (TU/e) from 2017 to 2018, and each log corresponds to one type of request types. Depending on the specific request type, employees can usually submit three types of documents which are travel declarations, travel permits and payment requests (Some event logs may not contain all document types). As described in the documentation, all documents follow a similar workflow, and the processes in 2017 and 2018 are different since 2017 is a pilot year. The information suggests that there is a potential process drift for the five logs sometime between the end of 2017 and the beginning of 2018.

We run our method on all five event logs without applying any noise filtering approaches. The description of each event log, window size used for drift detection, total execution time\footnote{The time includes converting the event log into event stream, forward detection and backward detection. The platform is the same as Section \ref{Section:5.4}.} and detection results are presented in Table \ref{bpic2020}.

\begin{table}[H]
\caption{Process drifts detection results on BPIC2020 data-sets.} \label{bpic2020}
\includegraphics[width=\textwidth]{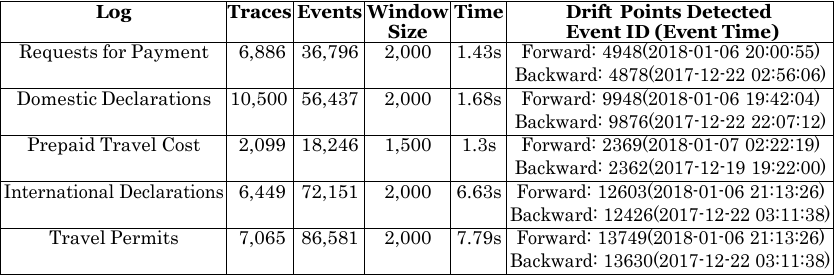}
\end{table}

As shown in Table.\ref{bpic2020}, the time for drift points is similar among the five logs. For each event log, the forward detection finds a drift point at the beginning of 2018 (new behaviours added), and the backward detection finds a drift at the end of 2017 (old behaviours removed). Besides, the number of events between the two drift points for each log is small (Although there is a relatively long time interval between the two drift points, we believe this is caused by the Christmas vacation). The results indicate that there is a process drift in each log sometime between the two detected drift points (at the end of 2017 or beginning of 2018) which involves both adding and removing of behaviours. The results conform to the documentation of the data-sets.

\begin{figure}[H]

\includegraphics[width=\textwidth]{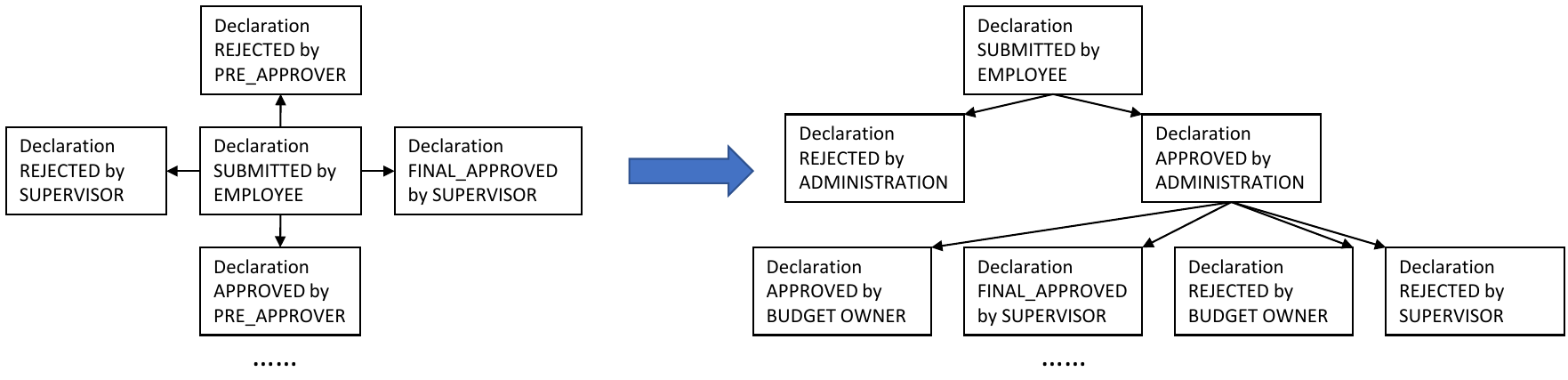}

\caption{Simple directly-follows graphs showing the process before (2017) and after (2018) the drift in the Domestic Declarations log.} \label{bpicdomestic}
\end{figure}

To further validate the results, we cut each event log into two sub-logs using the results of backward defections. We observe similar significant changes to all the five logs. Before the process drift, when a document is submitted, it can be sent to "pre-approvers" or supervisors for approval. After the process drift, the submission will be sent to the administration for approval, and if approved by the administration, it will be forwarded to the supervisor or budget owner for further steps. Fig.\ref{bpicdomestic} shows the process drift for the Domestic Declarations log.

Finally, it is worth mentioning that our method is efficient to detect process drifts. The time it takes to detect process drifts among all the five logs is within 10 seconds while three of the logs are completed within 2 seconds.

%\begin{table}[H]
%\centering
%\caption{Process drifts detection results on BPIC2020 data-sets}
%\label{bpic2020}
%\begin{tabular}{ |c|c|c|c|c|c| } 
%\hline
%
%\textbf{Log} & \textbf{Traces} & \textbf{Events} & \textbf{Window}& \textbf{Time} & %\textbf{Detected Drift Points} \\
%\hline
%
%Requests for Payment&6,886&36,796&2,000&1.43s&Forward: 4948(2018-01-06 20:00:55)\\
% & & & & &Backward: 4878(2017-12-22 02:56:06)\\
%\hline
%Domestic Declarations&10,500&56,437&2,000&1.68s&Forward: 9948(2018-01-06 19:42:04)\\
% & & & & &Backward: 9876(2017-12-22 22:07:12)\\
%\hline
%Prepaid Travel Cost&2,099&18,246&1,500&1.3s&Forward: 2369(2018-01-07 02:22:19)\\
% & & & & &Backward: 2362(2017-12-19 19:22:00)\\
%\hline
%International Declarations&6,449&72,151&2,000&6.63s&Forward: 12603(2018-01-06 21:13:26)\\
% & & & & &Backward: 12426(2017-12-22 03:11:38)\\
%\hline
%Travel Permits&7,065&86,581&2,000&7.79s&Forward: 13749(2018-01-06 21:13:26)\\
% & & & & &Backward: 13630(2017-12-22 03:11:38)\\
%\hline
%
%\end{tabular}
%\end{table}

\section{Conclusion}
\label{Section:7}
In this paper, we propose a new process drift detection method which can accurately locate the process drift points. If a valid drift can be identified, subsequent comparative analysis can be performed for process improvement. In addition, accurate process drift detection results can also be used as input for process drift characterization methods such as~\cite{Ostovar2017,Ostovar2020} to generate more accurate results. 

Different from previous work, our method does not rely on statistical tests to detect process drifts but applying statistical tests to differentiate between real process drift points and noise. The advantages of our method are as follows: First, The detection accuracy is high among event logs with different noise levels, and the high precision indicates the method returns very few false positives. Second, There is no need to define a noise filtering threshold, which reduces the need for background knowledge about the data. Lastly, The detection speed is reasonably fast.

It has to be noted that like other current window-based process drift detection algorithms, under different parameter settings, the detection results can still be different among different logs with different noise levels and with different process change types. In addition, process drifts which only contain branching frequency changes cannot be detected by the proposed method.

Future work includes the following aspects: First, we aim to propose a way to determine the window size automatically for different logs. Second, we plan to extend the work to characterize different drift types and provide comprehensive results. Finally, we aim to improve our work to suit online settings.
%
% ---- Bibliography ----
%
% BibTeX users should specify bibliography style 'splncs04'.
% References will then be sorted and formatted in the correct style.
%
% \bibliographystyle{splncs04}
% \bibliography{mybibliography}
%

\end{document}